\documentclass[aps,prl,twocolumn,showpacs,superscriptaddress,nobibnotes,groupedaddress]{revtex4}
\usepackage{amsmath, amsthm, amssymb}
\usepackage{graphicx}

\begin{document}

\title{Systemic risk in dynamical networks with stochastic failure criterion}

\author{B.~Podobnik}
\affiliation{Faculty of Civil Engineering, University of Rijeka, 51000 Rijeka, Croatia}
\affiliation{Zagreb School of Economics and Management, 10000 Zagreb, Croatia}
\affiliation{Faculty of Economics, University of Ljubljana, 1000 Ljubljana,  Slovenia}
\email{bp@phy.hr}

\author{D.~Horvatic}
\affiliation{Faculty of Science, University of Zagreb, 10000 Zagreb, Croatia}

\author{M.~Bertella}
\affiliation{Department of Economics, Sao Paulo State University (UNESP), Sao Paulo, Brazil}

\author{L.~Feng}
\affiliation{Department of Physics and Center for Computational Science and 
Engineering, National University of Singapore, Singapore 117456, Republic of Singapore}

\author{X.~Huang}
\affiliation{Department of Physics, Boston University, 590 Commonwealth Avenue, Boston, MA 02215, USA}

\author{B.~Li}
\affiliation{Department of Economics, Sao Paulo State University (UNESP), Sao Paulo, Brazil}
\affiliation{School of Physics Science and Engineering, Tongji University, Shanghai, 200092, P. R. China}

\begin{abstract}
Complex non-linear interactions between banks and assets
  we  model by  two  time-dependent  Erd\H{o}s Renyi network 
  models where each node, 
  representing bank, can invest either to a single asset ({model \bf I})
  or multiple assets ({model \bf II}). We use dynamical network approach
   to evaluate the collective financial failure---systemic risk---quantified 
   by the fraction of active nodes. The systemic risk can be calculated
  over any future time period, divided on sub-periods, where within each sub-period  banks may contiguously fail due to links to either (i) assets or (ii) other banks, controlled
    by two parameters, probability of internal failure $p$ and threshold $T_h$ (``solvency'' parameter). The systemic risk  non-linearly increases with  $p$ and  decreases 
    with average network degree
 faster when all assets are equally distributed
across banks than if assets are randomly distributed. 
  The more inactive banks each bank can sustain (smaller $T_h$), the
smaller the systemic risk---for some $T_h$ values in {\bf I} we report 
a discontinuity in  systemic risk. When contiguous spreading
 becomes stochastic (ii) controlled by probability $p_2$---a condition for
  the bank to be solvent (active) 
 is stochastic---the systemic risk decreases with decreasing $p_2$. 
We analyse asset allocation for the U.S. banks. 
\end{abstract}


\pacs{02.50.Ey,89.90.+N}

\maketitle

\section{Introduction}

  Phase transitions,  critical points, hystereses and regime shifts are basic
  blocks describing  the phase flipping
  of a complex dynamical system between two or more phases~\cite{May08}. 
  One of the systems having  these properties is the   financial system  
  that can be considered as a system  flipping
   over time between mainly stable phase, representing  good years, and mainly instable
  phase, representing bad years.  
 In the financial system, 
   the transition from mainly stable to mainly  instable phase 
   can be triggered  either by   an outside sudden event such as 
   a war or by a bankruptcy of a huge bank 
   where the financial contagion  can spread 
 due to interlinks between financial units. The nature of this contagion
  spreading   implies    that a network approach 
     can be the best suitable framework to describe not only  
     financial  contagion  but also financial crises~\cite{JPE1983,JPE2000,GaleAER,GolobAER,Gai10,Beale11,Guido13,May08,May10,May12,Lorenz,Acemoglu,Bisias,Schweitzer,Soramaki}. In  networks, just like in financial systems,  the  existing nodes 
 can  be rewired,  and new links and nodes can be added and removed  as time 
  elapses and  node's properties can change over time~\cite{Albert00,Vespignani01,Garlaschelli03,ParshaniPNAS,Cohen00,NC,Holme12}.

 In seminal work on network approach in finance, 
 Allen and Gale \cite{JPE2000} argued that  a more interelated network
 may help that  the losses of a disstressed bank are shared among more
creditors reducing the impact of negative shocks to each individual bank.
 Using a
 network model of epidemics where nework is characterized by  its degree
 distribution,  Gai and Kapadia \cite{Gai10} showed that a large rare shock may have
different consequences  depending on where it hits in the network and
what is the average degree of the network. 
 In contrast,  beyond a certain point, such
interconnections may serve as a mechanism for propagation of large shocks
leading to a more fragile financial system. 
 Beale {\it et al}~\cite{Beale11}  focused on vulnerability of
  financial system, precisely on 
 the friction of interest between  individual banks and entire economy.  
 The authors  
  investigated the relationship between  the risk taken by 
   individual banks and the 
    systemic risk  associate with multiple 
    bank failures. Arinaminpathy, Kapadia and May \cite{May12}
  reported how    
    imposing tougher capital requirements on larger banks than smaller
ones can  increase  the resilience of the financial  system.
 Elliot, Golob, and Jackson \cite{GolobAER} reported how integration (each
 organization becoming more dependent on its counterparties) and
 diversifications (each organization  interacting with a larger number of
 counterparties)  have different  effects on cascading failure. 
    Acemoglu, Ozdaglar, and A. 
Tahbaz-Salehi  ~\cite{Acemoglu} reported 
that financial contagion exhibits a phase transition as interbank
connectivity increases.   If  shocks are smaller than some threshold, more linked network enhance 
   the stability of the system. However, for shocks larger than the threshold,
    more linked network  facilitate financial contagion.   
  

\section{Results}
The global financial crisis has urged the need for analysing 
 systemic risk representing the collective financial failure~\cite{Lorenz,GolobAER,Gai10,GaleAER,May08,Beale11,Guido13,May10,May12}. 
 The majority of literature on systemic risk 
 is focused on how     the financial system responds to the failure of a
 single bank.  However, in real finance  many banks can fail inherently 
 either simultaneusly or at different  times. 
 Here, in order to  estimate the collective financial failure when multiple 
 initial failures are possible occuring presumably at various moments,  we 
   model complex non-linear interaction between banks and firms by 
  two variants of the 
  dynamical   Erd\H{o}s   Renyi network proposed in Ref.~\cite{Antonio14}, 
  where  
  nodes (banks)   contiguously fail due to links to both  (i) assets and (ii)  
      other banks,   and (iii) possibly recover.  In Ref.~\cite{Antonio14}, the 
   collective phenomena reported in a financial system---phase transitions,  
     critical points, hystereses and phase 
     flipping---have been   described  by the dynamical network approach.  
      It has been explained how the network, due to (ii) stochastic  
contiguous spread among the nodes, may 
lead to the spontaneous emergence of macroscopic phase-flipping phenomena.      
      In  our dynamical network    approach,
       bank $i$ can internally fail at any moment $t_i$, and once it fails,
    the contagion spreads at $t_{i+1}$ on $i$'s nearest neighbours, 
     at $t_{i+2}$ on $i$'s second neighbours, and so on. This approach 
   allows us   to  calculate  the systemic 
 risk at different future time horizons.  The probability 
    parameter,  controlling the
   macroscopic phase-flipping phenomena~\cite{Antonio14}, determines also that 
   the condition for a bank to be solvent (active) is not deterministic 
\cite{Gai10},   but stochastic.

   Qualitatively it is known 
    that more links between banks may reduce the risk of 
   contagion ~\cite{JPE2000}.    
   Recently a model of contagion in financial networks has been proposed~\cite{Gai10} where each bank $i$
  has interbank assets $A^B_{i}$, interbank liabilities $L^B_{i}$,
   deposits $D_i$  and illiquid assets $A^M_i$, and  
 the condition for the bank to be solvent (active) is 
    $ (1 - \phi) A^B_{i} + A^M_i - L^B_{i} - D_i > 0 $---a bank's assets must ecxeed its liabilities---where $\phi$
     is the fraction of inactive neighbouring banks.      
      To account 
     for possibility that many banks can fail at any moment---not only at initial moment---and to account for possibility to estimating bank risk for 
      different future time horizons, highly volatile financial 
      networks in this work we  model by a variant of the 
  dynamical Erd\H{o}s   Renyi  network~\cite{Antonio14}. We define  {\bf I}
   in the following: 
 \begin{itemize}
 \item[{(i)}]    
 nodes represent banks, and bank $i$ at time $t$ fails if 
  $ (1 - \phi_{t,i}) A^B_{i} + A^M_{t,i} - L^B_{i} - D_{t,i} \leq 0 $. 
 The previous time-dependent condition can be accomplished if (i) bank $i$
 internally  fails randomly and independently of other 
 nodes with probability of
 failure  $p$ (see Methods)---for each bank $i$, with probability $p$, 
  $A^M_{t,i} - D_{t,i}$ becomes negative and $i$  fails, regardless of interbank assets.    
 \item[{(ii)}] 
  Besides internally, in {\bf I}  bank $i$  can also fail  with 
 probability $p_2$ if  it has less 
  than $100 T_h\%$ active neighbours (where $i$ sets its interbank assets)~\cite{Antonio14,Pod13}.  
 If not stated differently, here we assume that   $p_2 = 1$, implying
 that if bank   $i$ has 
  $< 100 T_h\%$ active banks,       it deterministically fails. 
  The parameter 
   $T_h$ measures the robustness of  bank network---the smaller the parameter  $T_h$,
 the more robust the     bank network.  Note that 
     $T_h$ can be related with the criterion 
    for bank failures of Ref.~\cite{Gai10}. To this end, let's assume 
   that for each bank there is a linear
      dependence between asset $A^B_{i}$  and network degree $k$---e.g. 
      $A^B_{i} = k_i$.  
      Since the number of incoming and outgoing links is equal,  it holds $A^B_{i} = L^B_{i}$. 
     Expressing  
      $A^M_{i} $ and  $D_i $ at $t=0$   as proportional to $k_i(A^B_{i})$,
       when $i$ active,
       say
       $A^M_{t=0, i} = 0.6 k_i $ and $D_{t=0, i} = 0.3 k_i $, then  bank $i$ is 
       inactive if $\phi_{t,i} \ge 0.3\equiv \phi_h$ (if  
       at least $30\%$ neighbouring banks are inactive, or alternatively,
              if less than $70\%$ neighbouring banks are active). Thus, 
\begin{equation}    
       1 - T_h = \phi_h. 
\label{tphi}        
\end{equation}          
       \item[{(iii)}] 
       In   {\bf I} 
  after a time period $\tau$, the nodes recover from
  internal failure (see Methods).  In finance, this $\tau$ 
      is comparable with
      an average time a firm spends  in financial distress that
       is approximately two years for U.S. firms \cite{Eberhart}.  
  \end{itemize}

 \begin{figure}[b]
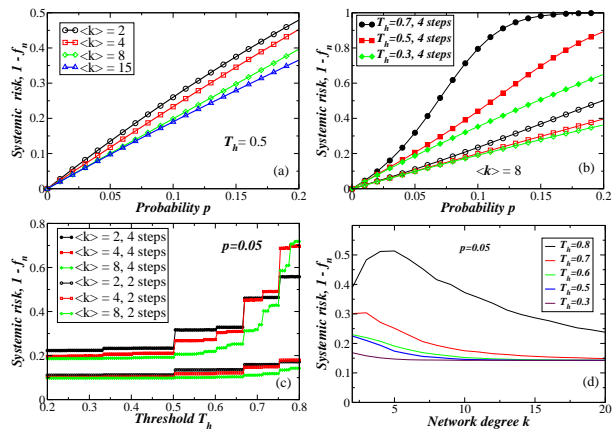

\centering \includegraphics[width=0.22\textwidth]{Fig1a.eps}
\centering \includegraphics[width=0.22\textwidth]{Fig1b.eps}\\
\centering \includegraphics[width=0.22\textwidth]{Fig1c.eps}
\centering \includegraphics[width=0.22\textwidth]{Fig1d.eps}
\caption
{   Dependence of network systemic risk of banks, $1 - f_n$
 (fraction of failed banks),  on network parameters,
 individual probability of bank failure $p$, threshold
      $T_h$, and network degree, $\langle k \rangle$. We use ER dynamical network 
       of Ref.~\cite{Antonio14}, where banks fail 
       independently of each other. 
 (a)-(b)  $1 - f_n$ increases with
  $p$ for varying $T_h$ and $\langle k \rangle$. 
  (c)  $1 - f_n$ non-continuously increases with  $T_h$.  
  (d) $1 - f_n$ decreases with  $\langle k \rangle$ for large values of 
  $\langle k \rangle$. 
}
\label{1}
\end{figure}

 To calculate how the systemic risk---among many different definitions \cite{Bisias}, 
  here defined as the fraction of failed
 banks---depends on the model parameters, first we present an 
 analytical result
  that holds for mean-field approximation, which is generally valid for a network with large   
  number of nodes and degrees.  
 If internal  ($X$) and  external ($Y$)  failures
  are  independent, Ref.~\cite{Antonio14} 
calculated  the probability, 
   $a = a(p,p_2,T_h) \equiv
  P(X \cup Y)$,  that a randomly chosen  bank node $i$   is  inactive
 $ a = p + p_2 (1 -p) \Sigma_{k}P(k) E(k,m,a)$. 
This probability is equal to 
  the fraction of inactive bank nodes, 
  $a = 1 - \langle f_n \rangle$. Here $P(k)$ is
   the degree distribution of the inter-bank links, and parameters $p$,
       $T_h$, and   
   $p_2$  are  explained  in   {\bf I}.   
 $E(k,m,a) \equiv 
  \Sigma_{j=0}^{m} a^{k-j} (1 - a)^{j} 
   {k \choose k -  j} $ is 
 the probability that node $i$'s neighbourhood is critically damaged             \cite{Antonio14}, 
 where $k$ is the number of links of node $i$, and   $m\equiv T_h k$. For  $m=1$  we provide
    an analytical form of  $a = 1 - \langle f_n \rangle$
 when $P(k)$ is the Poisson distribution 
  $P(k)= (\langle k\rangle)^k  e^{-\langle k\rangle } / k!$---we obtain  
 $ a=p+p_2(1-p)(1+\langle k\rangle-a \langle k\rangle)e^{\langle k\rangle(a-1)}$.
    
     Application of analytical results based on mean field approximation  
    in finance---where generally there are either small or
     moderately large number of  banks---is highly 
     limited since for these cases the 
  mean field holds only approximately~\cite{Antonio14}. 
   For these cases, in practice,
    numerical approach  helps us estimate how the systemic risk 
   quantitatively  depends on each model parameters and finally 
    enable us to, using
      regression, estimate  what the systemic risk is for a given set of
     empirically  estimated   parameters.

    For   dynamical Erd\H{o}s Renyi network {\bf I} with 1,000 banks, 
 in numerical simulations each of 10,000 runs  
 is used to estimate e.g. the systemic risk a year ahead expressed 
 as the fraction of   failed banks, $1 -f_n$.  
 Each run  we accomplish  
 in two time steps and each bank can internally fail in both time 
     steps with no recovery $(\tau > 2)$.  By  fixing parameters 
 $\langle k \rangle$ and  $T_h$,  in Fig.~(\ref{1})(a)-(b)   we  find that 
  the  bank
 systemic risk, $1 -f_n$,  increases non-linearly with 
  individual bank failure $p$. In Fig.~(\ref{1})(a) the systemic risk  
   decreases    with the average degree $\langle k \rangle$.  
 Thus, the larger the number of links 
  between banks, the smaller  the systemic   risk.  This result
    is in agreement with Ref.~\cite{JPE1983} where it was shown 
that networks with more links are less vulnerable
 than networks with a few links, since the fraction 
of the losses in one bank  is
transferred to more banks through interbank links. Note that our choice 
$\langle k \rangle =15$ 
 is due to  Ref.\cite{Soramaki}  reporting
  that  the average bank in the U.S. is linked  to $15$ others, however
   most banks has only few connections while a small number of huge banks  have thousands.   

 In Fig.~(\ref{1})(b)  we  report  that 
 the more failed neighbouring banks any bank can sustain 
 (the smaller $T_h$), the smaller the systemic 
 risk. In economics, the existence of threshold assumes that when 
 an organization's value (say asset minus debt) hits a failure threshold, 
 the organization discontinuosly can lose part of its value~\cite{GolobAER}.  
 Due to interdependencies among nodes, individual failues  may trigger 
  collective cascade of failures. To this end,  
  with increasing $T_h$
 the systemic risk in Fig.~(\ref{1})(c) exhibits
  a non-linear discontinuity  where $1 -f_n$  suddenly jumps
  at some critical points in $T_h$ such as $1/2, 2/3,...$, since link  values are integer numbers. Note that Ref.~\cite{Antonio14} reported 
 a discontinuity in the fraction of active nodes, $f_n$, when 
 increasing (decreasing)
  $p$ and $p_2$, together with the hysteresis property that is the 
  characteristic feature of a first-order phase 
  transition~\cite{Antonio14,Pod13}. As stated in Introduction,  recently 
  Ref.~\cite{Acemoglu}  reported phase transition as interbank links 
  increases.  Next,  
   Fig.~(\ref{1})(d) confirms again 
   that the 
   systemic risk in dynamical network approach
    decreases with  $\langle k \rangle$.

 For many parameter sets 
 ($\langle k \rangle$, $T_h$, $p$, $1 - f_n$) obtained from each two-period
  runs (where $\tau > 2$, thus no recovery), we perform linear regression analysis 
 $1 - f_n = \alpha + \alpha_p  p + \alpha_T T_h + \alpha_k k $, and obtain $\alpha=- 0.016\pm$ 0.001, 
  $\alpha_p = 2.20 \pm 0.03$, $\alpha_T = 0.067 \pm 0.005$,
  and  $\alpha_k = -0.0017 \pm 0.0002$. 
 The systemic risk significantly increases with $p$ and $T_h$, while decreases 
  with degree $\langle k \rangle$. Note that these values we obtained 
   using $p \in (0, 0.1)$, $T_h \in (0.2, 0.8)$, and 
   $\langle k \rangle \in (2, 20)$.

 Dynamical network approach reveals one more forecasting benefit. 
 It provides us
  with forecasting power for generally any future time horizon. 
 For the dynamical Erd\H{o}s Renyi network  {\bf I} in 
 Fig.~(\ref{1})(b)  we  show  the expected systemic risk  where
  each of 10,000 runs  
 is composed of  four steps. 
 If each  time step represents, say semi-year period, 
   than two-step  systemic risk represents our estimation for 
   systemic risk a year ahead, while four-step systemic risk
represents our estimation for 
   systemic risk two years ahead. As expected for the case with no recovery, four-step systemic risk
  is larger than two-step systemic risk. Similar result we  obtain
   in  Fig.~(\ref{1})(c) where for $\langle k \rangle = 8$ 
   we show that four-step systemic risk
   is substantially larger than  two-step systemic risk. Note that 
    these results we obtain under assumption that banks and assets,
     once failed, do not recover. Clearly, in more reliable 
     dynamical network approach one should also define how 
      assets and banks recover over time after,  for example, 
 the government     intervenes in the market.

\begin{figure}[b]

\centering \includegraphics[width=0.45\textwidth]{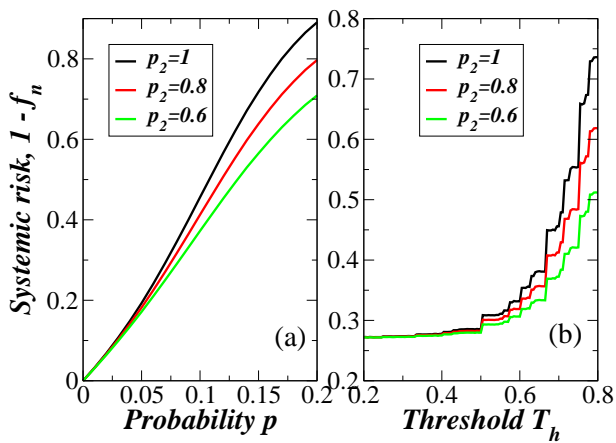}
\caption
{ For  fixed $p$ and (b) $T_h$, with decreasing parameter 
 $p_2$ (from 1 to 0), the systemic risk decreases. We set $\langle k \rangle = 8$
   and $T_h = 0.5$ in (a). 
}
\label{2}
\end{figure}

 Figure~(\ref{2}) shows that for fixed  $p$ and  $T_h$, the systemic risk decreases 
  with decreasing   parameter 
 $p_2$  (shown from 1 to 0.6), that is a reasonable result since $p_2=1$ implies that a bank deteministicly fails
   when a criterion for insolvency
    $ (1 - \phi_{t,i}) A^B_{i} + A^M_{t,i} - L^B_{i} - D_{t,i} \leq 0 $
     is fulfilled, while when $p_2 \ne  1$,
 there is some chance that the bank will not fail. 
 This unpredictability
 is something that really occurs in real market,
 since bankruptcy is not  deterministic event. However,
  stochasticity is not important only at microscopic firm (bank) level.
  Ref.~\cite{Antonio14} reported that introduction of 
  stochasticity leads to   
      emergence of macroscopic phase-flipping 
       between ``active'' and ``inactive'' macroscopic phases that
        are demonstrated  in Ref.~\cite{Pod13} for 
       ``expansion'' and ``recession'' phases in economy.

 In the previous  network  {\bf I} 
 it was assumed that each bank can independently internally fail. Next we 
 propose another network model {\bf II} (see Methods), where
  each bank can put its money
   not only in other banks, as in {\bf I}, but also 
  in different illiquid asset classes ${\cal A}_{t,i}^M$. Since different banks 
  can invest their money in equal  assets ${\cal A}_{t,i}^M$, 
   banks' failures are now not independent. 
   In contrast to banks' failures, ${\cal A}_{t,i}^M$'s failures are assumed to be
   independent   to each other.
    For simplicity,
  we define that firms  can affect banks, but not vice versa as in  \cite{Gai10,May12,Caccioli}. 


\begin{figure}[b]
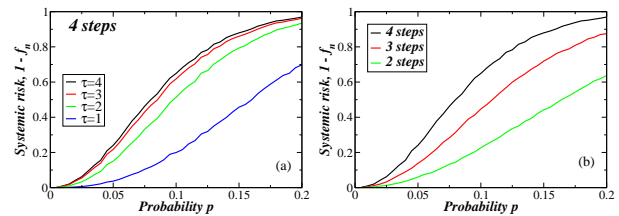

\centering \includegraphics[width=0.22\textwidth]{Fig3a.eps}
\centering \includegraphics[width=0.22\textwidth]{Fig3b.eps}

\caption
{ (a) Systemic risk for bank network  increases   
 with  time spent in failure $\tau$. (b) Systemic risk for different
   time scales. With increasing the time horizon, 
   the systemic risk 
    also increases. 
}
\label{3}
\end{figure}

\begin{figure}[b]

\centering \includegraphics[width=0.4\textwidth]{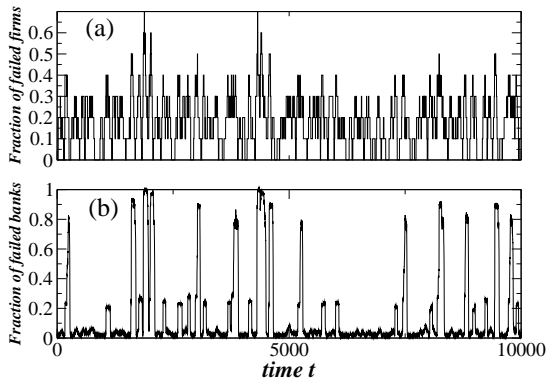}
\caption
{ Dependence between (a) firms and (b) banks  arises from 
 the dynamical  network model where firm failures 
  may affect all banks having investments in those failed firms.  
  We set $\tau = 50$, $p = 0.004$,  $p_2 = 0.8$, $\langle k \rangle=4$, 
  and   $T_h = 0.5$. Periods characterized by the largest spikes in the
   fraction 
    of failed firms coincide with the spikes in the fraction of  failed banks. 
 Both fractions of (a) failed  assets  
  and (b) failed  banks are increased where the fraction of failed banks are characterized
   by three states, one with virtually no failed nodes, and two states
    with different fractions of failed nodes.  
}
\label{4}
\end{figure}

 In the following simulations we include recovery process. 
  In {\bf II}  there are four time steps in our analysis, but this time we change
   the time period needed for recovery, $\tau$. 
  For (i)-(iii) ER dynamical network
   with  finite number of banks, $N_b = 1000$, and 
   assets, $N_f = 10$,  our   numerical simulations in Fig.~(\ref{3})(a)
 confirm that the systemic risk  calculated for bank
    network---defined  by the fraction of banks in failure---increases with the 
   individual  probability 
    of asset failure $p$, chosen to be equal  for each asset. We assume 
   as in Ref.~\cite{May12} that 
     both large and small banks hold the same number of asset classes, 
       $10$. We
     perform $10^5$ simulations in order to estimate expected systemic risk
     where each simulation itself we perform in  four 
      steps with $\tau$ meaning that once a bank or an asset is
       failed, it stays failed for the entire  period $\tau$. 
     Figures~(\ref{3}) reveals that this dynamical network  
     {\bf II}  exhibits highly  non-linear 
  properties.     By first fixing parameter 
    $T_h$,  in Fig.~(\ref{3})(a)  for  {\bf II}  we  find  that
     the systemic risk for bank network  increases   
 with   $\tau$. The longer time a bank stays in failure, the larger the systemic risk. 
  In Fig.~(\ref{3})(b) we show how dynamical approach enables us to estimate the systemic risk for different
   time horizons. As expected, with increasing the time horizon, 
   the systemic risk 
    also increases.

\begin{figure}[b]
\centering \includegraphics[width=0.4\textwidth]{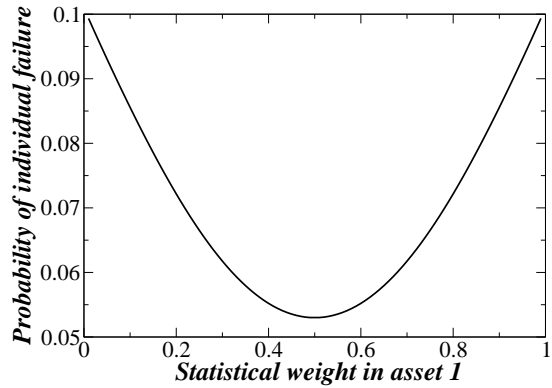}

\caption
{Smile-form in probability of individual bank failure with two assets with allocations 
 $W_1$ and  $W_2$. The smallest risk occurs as expected for 
 $W_1 = W_2 = 0.5$. We use $\tau = 2.5$.  
}
\label{5}
\end{figure}

\begin{figure}[b]
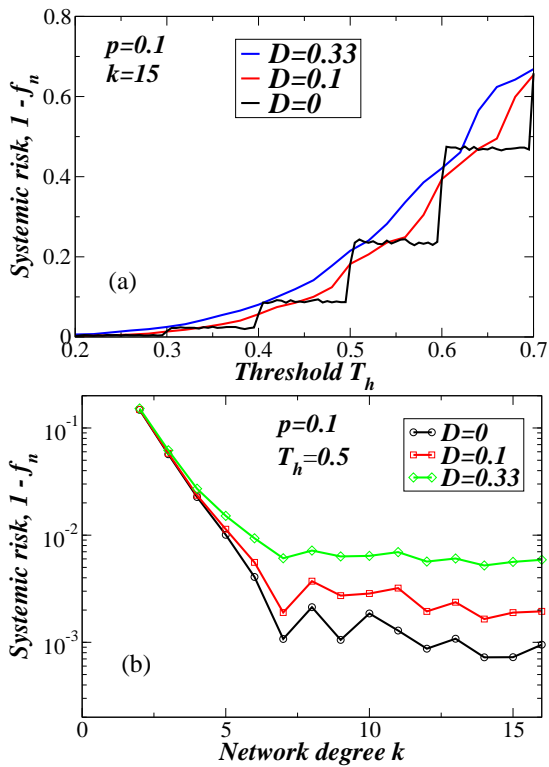

\centering \includegraphics[width=0.4\textwidth]{Fig6a.eps}

\centering \includegraphics[width=0.4\textwidth]{Fig6b.eps}
\caption
{Dependence of diversification of assets on
  systemic risk within dynamical network approach. (a) 
   Systemic risk  increases with $T_h$ for varying $D$
 value.    $D$  quantifies the level of asset
    allocation.  With increasing randomness in asset allocation,
  where $D$ increases from zero to $1/3$, the systemic risk 
   increases. We use that $p_2 = 1$ and $T_h = T'_h$ (see Methods). 
 (b)  Systemic risk decreases with average network degree 
  $\langle k \rangle$. 
}
\label{6}
\end{figure}

In order to analyse  
 how banks and assets   interact with each other over 
   time, we evolve simultaneously  both banks and assets. 
   In Fig.~(\ref{4}) we show how the fractions of failed $(N_f=10)$ assets and $(N_f=1000)$ 
   banks change over time for a given set of parameters,
   where recovery exists. Note  how the periods  
   when many assets are dysfunctional coincide  with the periods  
  when many banks are dysfunctional, in agreement with our 
  model assumption that asset failures 
  may affect bank failures in all
   banks having investments in the failed assets.

 Our network model {\bf II} is partially motivated by 
  Beale {\it et al}~\cite{Beale11}.  As  known in finance,
     each bank can reduce  its probability of failure  by diversifying 
     its risk \cite{Markowitz}.  However, when many banks
      diversify their risks in similar way, 
      the probability of multiple failure increases~\cite{Beale11}. 
 For the case with $N$ banks and $M$ assets,  \cite{Beale11}  defined  the total loss incurred by bank $i$ after one period $Y_i = \sum W_{ij} V_j$,  where the failure occurs if its total losses exceed a given threshold $\gamma_i$, i.e. $Y_i > \gamma_i$. Here, $W_{ij}$ denotes bank $i'$s allocation in asset $j$,  $V_j$  is the loss in asset $j$'s value taken from a student $t$ distribution, and $\gamma_i$ is a threshold.

  In  model   {\bf II} we use not $t$ distribution as in 
  Ref.~\cite{Beale11}, but a simpler Laplace 
 distribution
  and derive a  probability of bank failure with two assets with allocations
  $W_1$ and  $W_2$:
\begin{eqnarray*}
 P &=&\frac{ \exp(-\frac{\gamma}{W_2 \tau})}{4(1 + \frac{W_1}{W_2})} 
   + \frac{ \exp(-\frac{\gamma}{W_2 \tau})}{4(1 - \frac{W_1}{W_2})} 
   ( 1 -  \exp( - \frac{(1 - \frac{W_1}{W_2})\gamma}{W_1\tau}))   \nonumber \\
 &+& 
  \frac{ \exp(-\frac{\gamma}{W_1 \tau})}{2}  +
\frac{\exp(\frac{\gamma}{W_2 \tau})}{4(1 + \frac{W_1}{W_2})} 
  \exp( -\frac{ (1 + \frac{W_1}{W_2})\gamma}{W_1\tau}).
\end{eqnarray*}
 This expression, numerically tested in Fig.~(\ref{5}),  gives the same smile-form for the probability of
  bank failure
  as numerically found in 
  Ref.~\cite{Beale11}. 
   Next we analyse how diversification of asset
   allocations affects  systemic risk within the dynamical network approach.  
   To this end, Ref.~\cite{Beale11} proposed a measure to estimate the
    level of asset
    allocation 
\begin{eqnarray}
  D = \frac{1}{2 N (N-1)} \sum_{i=1}^{N} \sum_{j=1}^{N}
  \sum_{l=1}^{M} |W_{i,l} - W_{j,l}| 
 \label{D}
\end{eqnarray} 
     in order to quantify the average of the distances between each pair 
      of banks' asset allocations, where $D=0$ if each bank 
      invests equally in each asset, thus if $W_{i,j}=1/M$ for each
      $i$ and $j$. Thus, when many banks decide to invest with similar portfolios,
      they may increase the chance  to fail simultaneously.     
  In  Fig.~(\ref{6})(a) within dynamical network approach, 
   we obtain that the lowest systemic risk  occurs when all assets are 
    equally distributed across banks, 
     that is in agreement with the result  obtained in  
  Ref.~\cite{Beale11} where banks do not affect each other, and
   systemic risk is  defined as expected number of failures. 
   With increasing randomness in asset allocation,
  where $D$ increases from zero to $1/3$, the systemic risk 
   also increases. As a new result arising from the network approach,  
   in   Fig.~(\ref{6})(b) we obtain that 
    the systemic risk decreases with average degree
     $\langle k \rangle$ faster when asset allocation
      is more homogeneous.
    
 To see how well the real market can withstand systemic risk, we exam the  level of diversification of asset allocation for the U.S.
  commercial banks. Here, based on 
 64,289 commercial banks' balance sheet, we analyze their loan allocations in 10 different sectors~\cite{Huang13}: Loans for Construction and Land Development
Loans Secured by farmland, 
Mortgages Secured by 1-4 Family residential properties, 
Loans Secured by multifamily ($> 5$) residential properties, 
Loans Secured by nonfarm nonresidential properties, 
Agricultural Loans, 
Held-to-Maturity securities, 
Commercial and industrial loans, others, 
Available-for-sale securities, 
Loans to Individuals. 
 The first five sectors are under the asset class of Real Estate which is the  biggest part of the banks' loans allocations. 
Hence we decompose them to smaller sub-sectors. The data are collected from 1/1/1976 - 12/31/2008  and a considerable fraction of the banks have disappeared during the period. We use Eq.~(\ref{D}) to calculate the average diversification of the commercial banks, and find  $D=0.51$, a larger value  than $D=0.33$, obtained for random
  diversification.  It indicates that in real market, diversification of banks puts the systems at a  riskier regime than random diversification.
 
 \section{Discussion}
  Generally, the systemic  risk risk may arise either  because banks 
    diversify their risks in firms in similar ways \cite{Beale11} or because banks are 
linked within a bank network and thus failures can spread  contiguously\cite{JPE1983,JPE2000,GaleAER}.
 For  the dynamical (time-dependent) network 
 \cite{Antonio14} where nodes (i) inherently
fail, (ii) contiguously fail, and (iii) recover, when applied to finance, 
 the stochastic
contiguous spreading controlled by parameter $p_2$  corresponds to a  stochastic
 condition for the bank
to be solvent (active in network terms).  We demonstrated how the  systemic 
   risk in the dynamical (time-dependent) network depends on dynamical network parameters: 
   probability of individual asset failure $p$, 
 bank vulnerability parameter   $T_h$, and average network degree
  $\langle k \rangle$. 
 We  found  that the more inactive banks each bank can sustain  
  ($T_h$ closer to zero), the
smaller the systemic risk.  For fixed $p$ and $T_h$, the systemic risk decreases
 with decreasing parameter $p_2$ (increasing stochasticity).  
 This result is reasonable because  when a criterion for a bank's insolvency is met
   and $p_2 \ne  1$,
 there is some chance that the bank will not fail.  
 In practice, the parameter $T_h$ can be controlled by the 
 government or  banks. Our analysis showed that having $T_h \lesssim  0.4$ enables 
 the system to function with low risk for a large set of parameters $p$ and
  $\langle k \rangle$. We demonstrated how the dynamical network approach enables
 us to estimate systemic risk for different future time scales. We also 
  showed how the systemic risk  depends on  the level
of asset allocation. 
  The dynamical network approach enables to estimate 
  systemic risk in a wide range of 
  financial systems from pension, mutual, and hedge funds to banking system.

\section{Methods}

\begin{figure}[b]
\centering \includegraphics[width=0.4\textwidth]{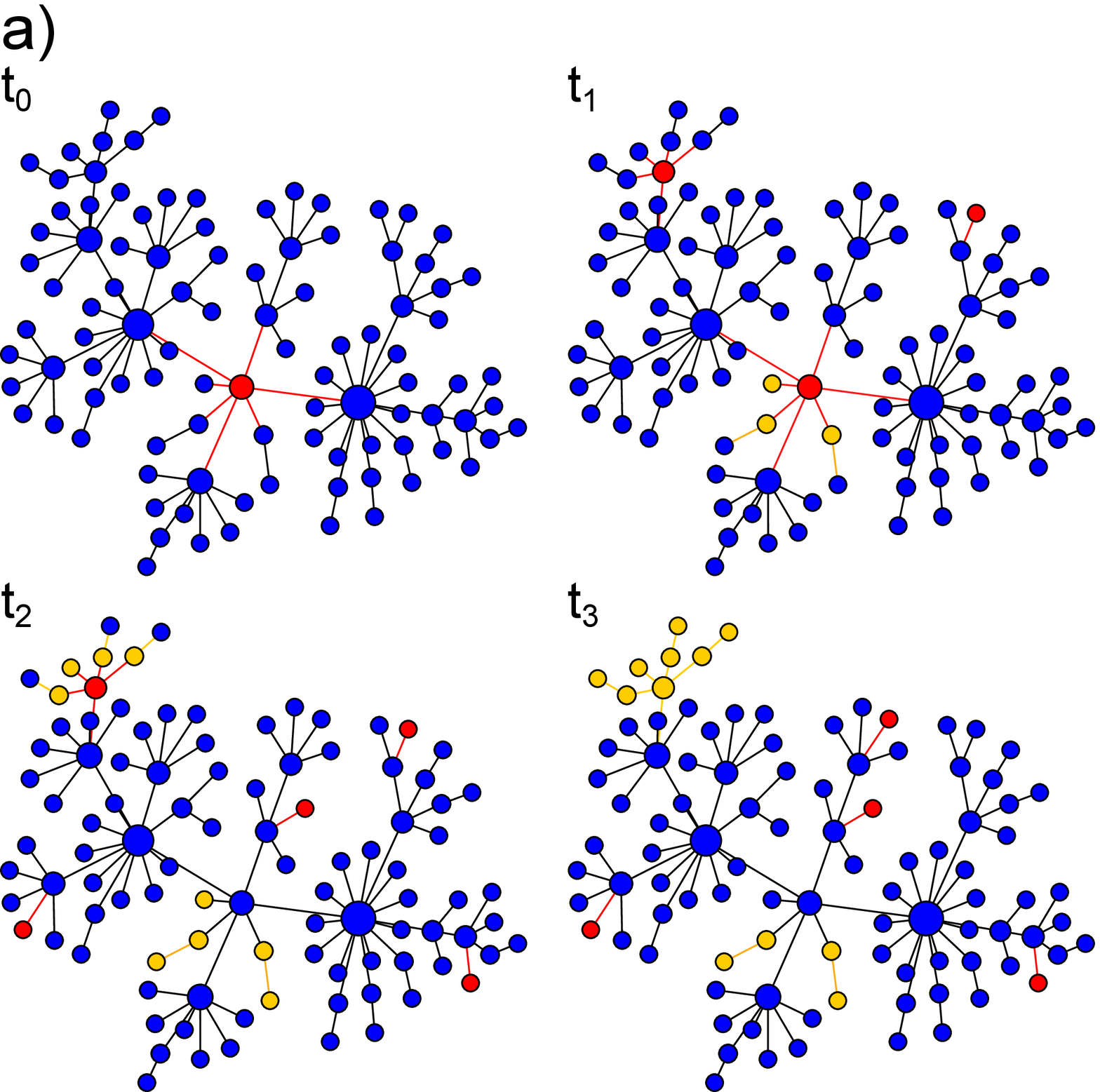}
\vspace{-5mm}
\centering \includegraphics[width=0.4\textwidth]{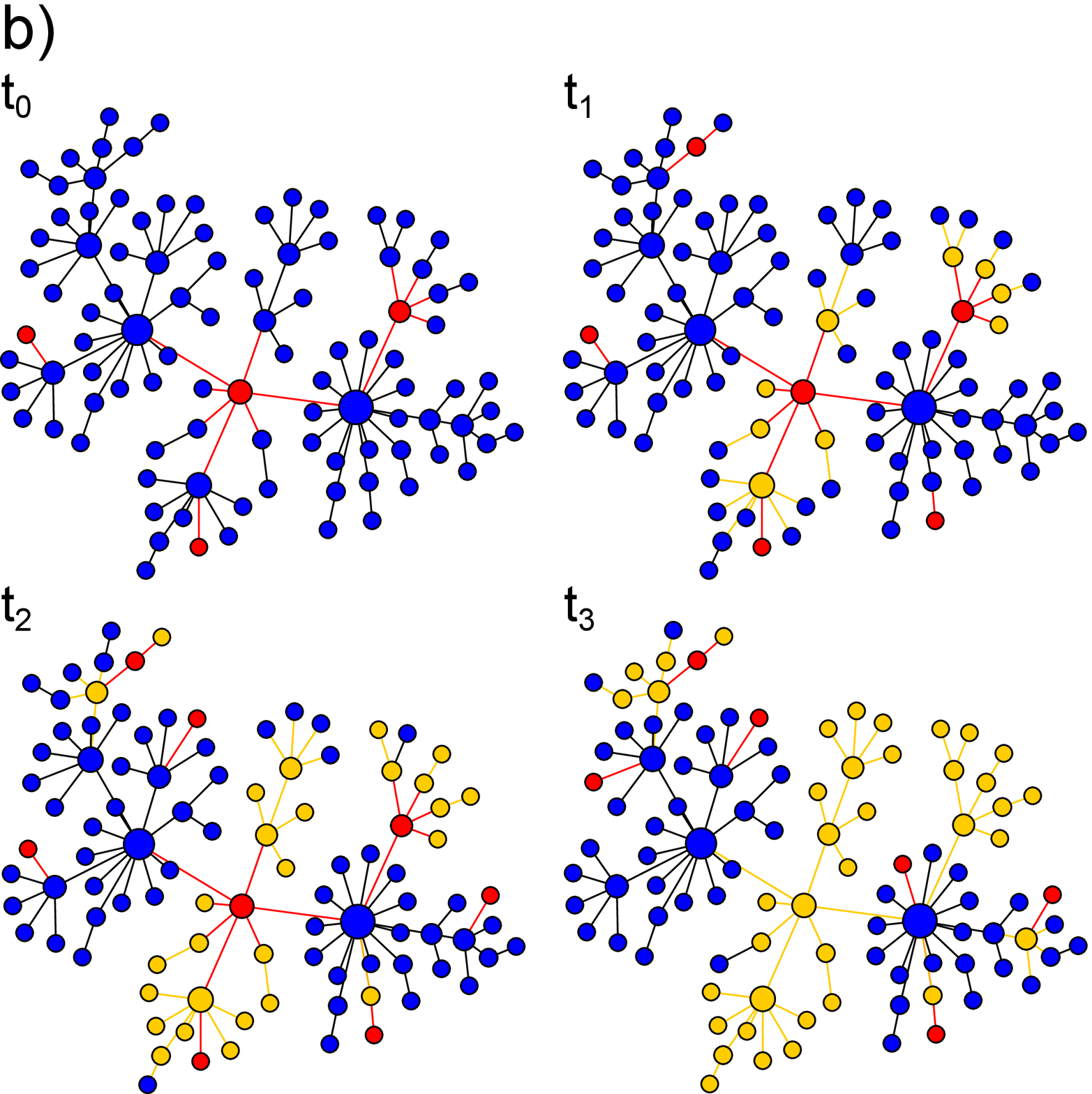}
\caption
{Contiguous spread in model I.  (a) Deterministic  solvency criterion 
($p_2 = 1$).
For steps $t = 0, 1, 2$, and $3$ shown is a network
 for $\tau = 2$. 
 Active nodes are blue, internally inactive nodes are 
 red and externally inactive nodes are denoted by orange. Nodes can become 
  internally inactive at any moment where congigion is spread 
  each step towards the nearest neighbours. (b) Stochastic solvency criterion
   $p_2 \ne  1$).  
   For steps $t = 0, 1, 2$, and $3$ shown is a network
 for $\tau = 3$ and $p_2 = 0.8$. 
}
\label{7}
\end{figure}

 We defined the network model {\bf I} in Results and here we explaine
  the model graphically. 
  In Fig.~(\ref{7})(a)  we choose $\tau = 2$, $T_h = 0.5$, and 
   $p_2 = 1$, where the last choice for   $p_2$ implies
    that  the solvency criterion is   
    deterministic.  Due to dynamical time-dependent network approach, 
     in Fig.~(\ref{7})(a)
     we see how 
      different nodes become   internally failed   
  at different times. Due to the choice for the value of   $T_h$, 
   the financial contigion is not spread  to the nearest hubs
  (note that a bank deterministically fails if 
      it has less   than $100 T_h\%$ active neighbours).  
      In Fig.~(\ref{7})(b)  we choose $\tau = 3$ and  
   $p_2 = 0.8$ where the last choice for  $p_2$ implies that 
    the solvency criterion is now 
  stochastic. Now we choose the larger   $T_h$ than in (a), 
  $T_h = 0.8$, meaning that each bank is more dependent to failures of its neigbours than in (a).  
 Now we see in Fig.~(\ref{7})(b)  how the financial spreading  is more devastating then in case (a). 
  Note that, due to stochasticity in  the solvency criterion,  
  some nodes which would be externally inactive due to deterministic criterion in (a)   are now in (b) externally active.

 We define the network model {\bf II}   in the following:  
 \begin{itemize}
 \item[{(i)}]    
At each time $t$,  each of $N_f$ assets ${\cal A}_{t,i}^M$ 
  can independently fail 
  where the probability $p$ of failure for each  ${\cal A}_{t,i}^M$ is equal 
  and is taken from a Laplace 
  (double) exponential   distribution, ${\cal L}(x)$.
   We define that  asset ${\cal A}_{t,i}^M$ fails if the Laplace variable $x$ 
  is smaller than some threshold, 
   $P_h$---thus,  $p = \int_{-\infty}^{P_h} dx {\cal L}(x)$. 
For simplicity, assets  do not influence each other. 
Once  asset  ${\cal A}_{t,i}^M$ fails,  
   it stays in failure with no possibility for recovery. 
 \item[{(ii)}] 
 Each of the $N_b$ nodes representing  banks, for 
 simplicity reason, 
 has the same asset and liability values as in example in  
 {\bf I}---so, bank $j$ at time $t=0$ has $A^B_{j} = k_j$, 
  $A^M_{t=0,j} = 0.6 k_j $ and $D_j = 0.3 k_j $ (here 
  $D_j$ does not change in time), 
  but the results are robust to different allocations. 
 However, this time the total illiquid asset $A^M_{t=0,j}$ is allocated 
 across   $N_f$ assets.   
  Precisely, 
 at initial time $t=0$ every bank $j$ determines  how much money
  $W_{j,i}$  to invest in
   asset ${\cal A}_{i}^M$, where $W_{j,i}$  is taken randomly 
  from the homogeneous distribution after proper  normalization, where
  clearly for each $j$, $\sum_{i=1}^{N_f} W_{j,i} = A^M_{t=0,j}$,
   where the sum runs over all assets.  Since at each moment 
  assets can be either active or inactive
  (failed),  at each  $t$,   bank $j$ has the total
 illiquid asset    equal to 
 $A^M_{t,j} = \sum_{i=1}^{N_f} W_{j,i} \delta_i(t) \le A^M_{0,j}$, 
 where $\delta_i(t)$ at time $t$ can be either
  one or zero depending whether asset ${\cal A}_{i}^M$ is active or inactive. 
  The more  robust the bank, the larger  the fraction of failed 
  assets the bank can 
  sustain without getting failed. 
  Here we  define that bank $j$'s node becomes  internally contiguously 
    failed  due to links to  assets  
  if $A_{t,j}^M \le T'_h A_{t=0,j}^M $, where  $T'_h $ is 
  the given threshold. Bank $j$
   internally fails if  $A^M_{t,j}  - D_{j} <  0$ and from our 
  choice for $A^M_{t=0,j}=0.6k_j$  and $D_{j} =0.3 k_j $ the failure 
  occurs when $T'_h = 0.5$.  We define bank node $j$'s internal
      failure state by  spin $|s_j \rangle$. 
  \item[{(iii)}] 
In our financial network,
     a  bank can fail  either due to illiquid asset' failures or 
    due to banks' failures.
   Here the concept of a network is used to model
 interbank lending and to study the
phenomena of financial stability and contagion~\cite{JPE1983,JPE2000,GaleAER}. At initial time, banks  create links among themselves through exchanging deposits to insure themselves 
against contagion~\cite{JPE2000}.  To this end, 
 external failure  state of bank node $j$  denoted by 
       spin $|S_j \rangle$ 
   is $| 0\rangle$ (during a time $\tau' = 1$)
    with  probability 
     $p_2$ if less than $100 \cdot T_h ~\%$ of $j$'s neighbouring links are      active~\cite{Antonio14}. 
      Bank node $j$---described
       by the  two-spin state
  $| s_i,S_i  \rangle$---is active only if both spins are  1, i.e,
   $| s_i,S_i \rangle =  | 1, 1 \rangle$. Links between banks are 
    bidirectional. Having two spins, representing a 
    financial health of a bank,  assumes
  that a bank fails either if it made bad investments in firms or in 
  other banks---alternatively if it is surrounded by bad neighbours.
  We assume  $A^M_{0,j}=0.6k_j$, $D_{j} =0.3k_j $ (the same choices as in 
  {\bf I}), and 
   $A^B_{j} = k_j$.  For this choice of allocations, if we assume
    that all illiquid assets where $j$ invested
     are active, we obtain  $T_h = 0.7$ $(\phi_h = 0.3)$
      (see Eq.~(\ref{tphi})).  
     Suppose that at time $t$ some illiquid assets ($20\%$) are inactive 
     and $A^M_{t,j} = \sum_{i=1}^{N_f} W_{j,i} \delta_i(t) = 0.8 A^M_{0,j}
     = 0.48 k_j $. 
     From 
     insolvency criterion
      $ (1 - \phi_{t,i}) A^B_{i} + A^M_{t,i} - L^B_{i} - D_{t,i} \leq 0 $ 
      we obtain $\phi_{t,i} \geq  0.18 \equiv \phi_h$. Hence, 
      the larger the  fraction of inactive illiquid assets, 
      the smaller the fraction of  neighbouring banks 
      required to cause the external failure.

 \end{itemize}

\begin{thebibliography}{99}




 \bibitem{May08}
R. M. May,  S. A.    Levin,  and G.  Sugihara, 
  Nature\/ {\bf 451}, 893 (2008).


\bibitem{JPE1983} 
 D. W. Diamond and P. H. Dybvig,  
 Journal of Political Economy {\bf 91}, 401 (1983). 

\bibitem{JPE2000} 
F. Allen and D. Gale, 
Journal of Political Economy {\bf 108}, 1 (2000).

\bibitem{GaleAER}   
 D. M.  Gale and S. Kariv, 
 American Economic Review {\bf 97}, 99 (2007). 

\bibitem{Soramaki}  
K. Soramaki   {\it et al},  
  Physica A\/ {\bf  379}, 317 (2007). 



 \bibitem{Schweitzer}  F. Schweitzer  {\it et al}, 
  Science\/ {\bf 325}, 422 (2009).

\bibitem{Lorenz} 
J. Lorenz, S. Battiston, and F. Schweitzer,  
 Eur. Phys. Journal B\/ {\bf 71}, 441 (2009). 



 \bibitem{May10}  R. M. May and N. Arinaminpathy, 
   J. R. Soc. Interface\/ {\bf 7}, 823 (2010).



\bibitem{Gai10}
P. Gai and S.  Kapadia,  
 Proceedings of the Royal Society of London  A\/ {\bf 466}, 2401 (2010). 

\bibitem{Beale11}  N. Beale {\it et al},  
  Proc. Natl. Acad. Sci. USA {\bf 108}, 12647 (2011).


 \bibitem{May12} 
N. Arinaminpathy,  S.  Kapadia,  and R. M. May, 
 Proc. Natl. Acad. Sci. USA {\bf 109}, 18338 (2012).


 \bibitem{Guido13} D. Delpini  {\it et al},  
  Sci. Rep.\/ {\bf 3}, 1626 (2013).






\bibitem{Acemoglu} 
D. Acemoglu, A. Ozdaglar, and A. Tahbaz-Salehi (2012), 
http://ssrn.com/abstract=2207439. 

\bibitem{Bisias}
D. Bisias, M. D. Flood, A. W. Lo,  
S. Valavanis,
    Annual Review of Financial Economics\/ {\bf 4}, 255 (2012).





\bibitem{GolobAER}   
 M. Elliott, B. Golub, and  M. O. Jackson, forthcoming in the
{American Economic Review}. 







\bibitem{Albert00} R. Albert, H. Jeong,  and A. -L. Barab{\'a}si.  
  Nature {\bf 406}, 378 (2000).

\bibitem{Vespignani01}
R. Pastor-Satorras and A. Vespignani,
 Phys. Rev. Lett. {\bf 86}, 3200 (2001).

\bibitem{Garlaschelli03} D. Garlaschelli, G. Caldarelli,  and 
 L.  Pietronero,  
  Nature {\bf 423}, 165 (2003).

\bibitem{ParshaniPNAS}  R. Parshani, S. V.  Buldyrev, and S. Havlin,
 Proc. Natl. Acad. Sci. USA {\bf 108}, 1007 (2011).

\bibitem{Cohen00}
R. Cohen, K. Erez, D. ben-Avraham, S. Havlin, 
 Phys. Rev. Lett. {\bf 85}, 4626 (2000). 

\bibitem{NC}
 A. Bashan {\it et al}, 
 Nature Communications {\bf 3}, 702 (2012).



\bibitem{Holme12}
  P. Holme and J.  Saramaki, 
   Phys. Reports {\bf 519}, 97 (2012). 


\bibitem{Antonio14}
 A.  Majdandzic  {\it et al.},  
  Nature Physics {\bf 10}, 34 (2014). 


\bibitem{Pod13}  B. Podobnik {\it et al}, 
 arXiv:1401.7450 (to appear in PRE). 




\bibitem{Eberhart}      A. C. Eberhart,
    W. T. Moore, and 
    R. L. Roenfeldt, 
 J. Finance {\bf 45}, 1457 (1990).

\bibitem{Caccioli}
F. Caccioli, T. A. Catanach, and J. D. Farmer, 
 	arXiv:1109.1213.  



 \bibitem{Markowitz} H. Markowitz,  
  J. Finance\/ {\bf 7}, 77 (1952).

 

 \bibitem{Huang13} X. Huang, I. Vodenska, S. Havlin, and H. E. Stanley,  
  Scientific Reports\/ {\bf 3}, 1219 (2013).



\end{thebibliography}
\end{document}